\documentclass[aps,prc,reprint,superscriptaddress,nofootinbib]{revtex4-2}

\usepackage{amsmath}
\usepackage{amssymb}
\usepackage{graphicx}
\usepackage{dcolumn}
\usepackage{bm}
\usepackage{braket}
\usepackage{mathrsfs}
\usepackage{slashed}

\begin{document}


\title{
Entanglement generation in few-nucleon scattering
}

\author{Dong Bai}
\email{dbai@hhu.edu.cn}
\affiliation{College of Science, Hohai University, Nanjing 211100, China
}%

\author{Zhongzhou Ren}
\email{zren@tongji.edu.cn}
\affiliation{School of Physics Science and Engineering, Tongji University, Shanghai 200092, China}%
\affiliation{Key Laboratory of Advanced Micro-Structure Materials, Ministry of Education, Shanghai 200092, China}


\begin{abstract}

Inspired by a recent Letter [S.~R.~Beane \emph{et al.}, Phys.\ Rev.\ Lett.\ {\bf 122}, 102001(2019)],
the entanglement generated in the elastic $S$-wave scattering of $p+{}^{3}\text{He}$ and $n+{}^3\text{H}$ is studied,
where the proton, neutron, ${}^{3}$He, and ${}^3$H are all regarded as qubits.
To deal with the Coulomb interaction between the proton and ${}^{3}$He, we derive the entanglement power,  
a physical quantity that measures the average entanglement generated by a scattering process, 
for charged qubits within the screening method.
The entanglement power in the aforementioned two few-nucleon scatterings is found to be generally much smaller than that in the $S$-wave $n+p$ scattering at low energies,
with the corresponding cluster effective field theories possessing an enhanced approximate $\text{SU}(2)_1\otimes\text{SU}(2)_{2}$  symmetry at leading order.
Our study suggests that the entanglement generation capacities of effective interactions between nucleons and light nuclei 
could be more suppressed than realistic nucleon-nucleon interactions
 at low energies.

\end{abstract}

\maketitle


\section{Introduction}

Entanglement is a deep property of quantum mechanics, 
referring to the fact
that the quantum state of a composite system cannot always be decomposed into the tensor product of the quantum states of its constituents.
Nowadays, it is widely recognized as one of the sharpest probes to distinguish quantum mechanics from classical mechanics (even with local hidden variables) \cite{Nielsen:2010}.  

In the past, most efforts in nuclear physics were devoted to solving nuclear Hamiltonians with different methods
and calculating structural and reaction observables across the nuclide chart.
In comparison, the entanglement aspects of nuclear physics have been explored only recently.
In Refs.~\cite{Legeza:2015fja,Gorton:2018,Johnson:2019,Kruppa:2020rfa,Kovacs:2021yme,Kwasniewicz:2013cqa,Kwasniewicz:2017dbc,Kwasniewicz:2016,Johnson:2022mzk},
the entanglement properties of valence nucleons are explored in nuclear shell models.
In Refs.~\cite{Kanada-Enyo:2015ncq,Kanada-Enyo:2015kyo},
the entanglement entropy is adopted to distinguish between
 the Brink wave function and the Tohsaki-Horiuchi-Schuck-R\"opke wave function, two variational bases in microscopic cluster models.
In Ref.~\cite{Robin:2020aeh}, the entanglement properties of \emph{ab initio} wave functions are studied
for the light nuclei ${}^{4}$He and ${}^{6}$He,
suggesting that they can be the useful guidance for developing efficient \emph{ab initio} algorithms.
In Ref.~\cite{Jafarizadeh:2022kcq}, the entanglement entropy is adopted to study quantum phase transition in nuclei.
In Ref.~\cite{Pazy:2022mmg}, the entanglement entropy is calculated for the short-range correlated nucleon-nucleon pairs in nuclei
and is found to be linearly dependent on the mass number.
In Ref.~\cite{Tichai:2022bxr}, the orbital entanglement is investigated within a new hybrid framework called valence-space density matrix renormalization group.

In order to fully explore the entanglement aspects of nuclear physics,
it is crucial to understand the entanglement generation capacities of nuclear interactions.
%
The entanglement power measures the average entanglement generated by the $S$ matrix from arbitrary unentangled in-states  \cite{Zanardi:2001zza}.
By calculating the entanglement power for nuclear scatterings,
one can get useful information on the entanglement generation capacities of nuclear interactions.
The realistic nucleon-nucleon interactions (e.g., AV18 \cite{Wiringa:1994wb} and Idaho-$\text{N}^3\text{LO}$ \cite{Entem:2003ft}) have a number of spin operators
to manipulate the spin wave functions of nucleons.
It is thus natural to expect that they are good at generating entanglement in spin space.
In Ref.~\cite{Beane:2018oxh}, Beane \emph{et al.}\ study
the spin-space entanglement generation in the $S$-wave $n+p$ scattering,
with the neutron $n=(n_\uparrow,n_\downarrow)^T$ and the proton $p=(p_\uparrow,p_\downarrow)^T$ regarded as two distinguishable qubits. 
Indeed, it was found that the entanglement power 
approaches its maximal value twice in the $S$-wave $n+p$ scattering for relative momenta $p\leq80$ MeV (see also Fig.~\ref{EPp3He} in Sec.\ \ref{NR})  \cite{Beane:2018oxh}.
It is argued that
the approximate Wigner SU(4) symmetry that emerged in low-energy nuclear physics \cite{Wigner:1936dx,Mehen:1999qs} 
is closely related to minimizing the entanglement power \cite{Beane:2018oxh}.
Later on, this connection between entanglement minimization and symmetry enhancement was polished in Ref.~\cite{Low:2021ufv}.

Up to now, our knowledge on the entanglement generation capacities of nuclear interactions has been limited to hadron-hadron interactions \cite{Beane:2018oxh,Beane:2021zvo,Liu:2022grf}.
This motivates the authors to study the entanglement generation in the low-energy $S$-wave $p+{}^{3}\text{He}$
and $n+{}^3\text{H}$ scatterings,
which are examples of quantum collisions between hadrons and light nuclei,
and provide information on the entanglement generation capacities of effective interactions between nucleons and light nuclei. 
Like nucleons, ${}^{3}$He and ${}^3$H are spin-1/2 particles and thus can be regarded as qubits below their disintegration thresholds.
Compared with Ref.~\cite{Beane:2018oxh},
we study
the impact of the Coulomb interaction on entanglement generation
with the help of the screening method where the Coulomb interaction is cut off at large distances.
As far as we know, such a study has not been carried out explicitly in the literature yet.
The $p+{}^{3}\text{He}$ and $n+{}^3\text{H}$ scatterings have attracted much attention recently in the \emph{ab initio} community \cite{Lazauskas:2004uq,Deltuva:2006sz,Deltuva:2007xv,Viviani:2011ax,Viviani:2013wra,Viviani:2016cww,Viviani:2020tzc,Flores:2022foz}.
They are also investigated in pionless effective field theories (EFTs) \cite{Kirscher:2011uc,Contessi:2022vhn,Schafer:2022hzo}.
Besides the aforementioned two scattering processes,
the $p+{}^{3}\text{H}$
and $n+{}^3\text{He}$ scatterings could be regarded as qubit-qubit scatterings as well.
However, they turn out to be more complicated than the $p+{}^{3}\text{He}$
and $n+{}^3\text{H}$ scatterings.
For example,
the charge exchange channel $n({}^{3}\text{He},{}^{3}\text{H})p$ is open in the $n+{}^3\text{He}$ scattering from the beginning, making
the effective range parameters complex valued \cite{Hofmann:2003av}.
We postpone the comprehensive studies on these two scatterings to future publications.

This paper is organized as follows: In Sec.\ \ref{TF},
the mathematical properties of the entanglement power are investigated for elastic scatterings between two charged qubits,
which are relevant to the $p+{}^{3}\text{He}$ scattering.
The results for neutral qubits could be obtained easily from the charged ones
by taking all the charges to be zero (or equivalently, taking the fine structure constant to be zero).
The natural units $\hbar=c=1$ are adopted in theoretical derivations.
In Sec.\ \ref{NR},
the spin-space entanglement generation
is studied in detail for the low-energy $S$-wave $p+{}^{3}\text{He}$ and $n+{}^3\text{H}$ scatterings, along with the symmetry enhancement in the corresponding cluster EFTs.
Section \ref{Concl} summarizes and concludes.

\section{Entanglement Power for Charged Qubits}
\label{TF}

\begin{figure}

  \includegraphics[width=0.9\linewidth]{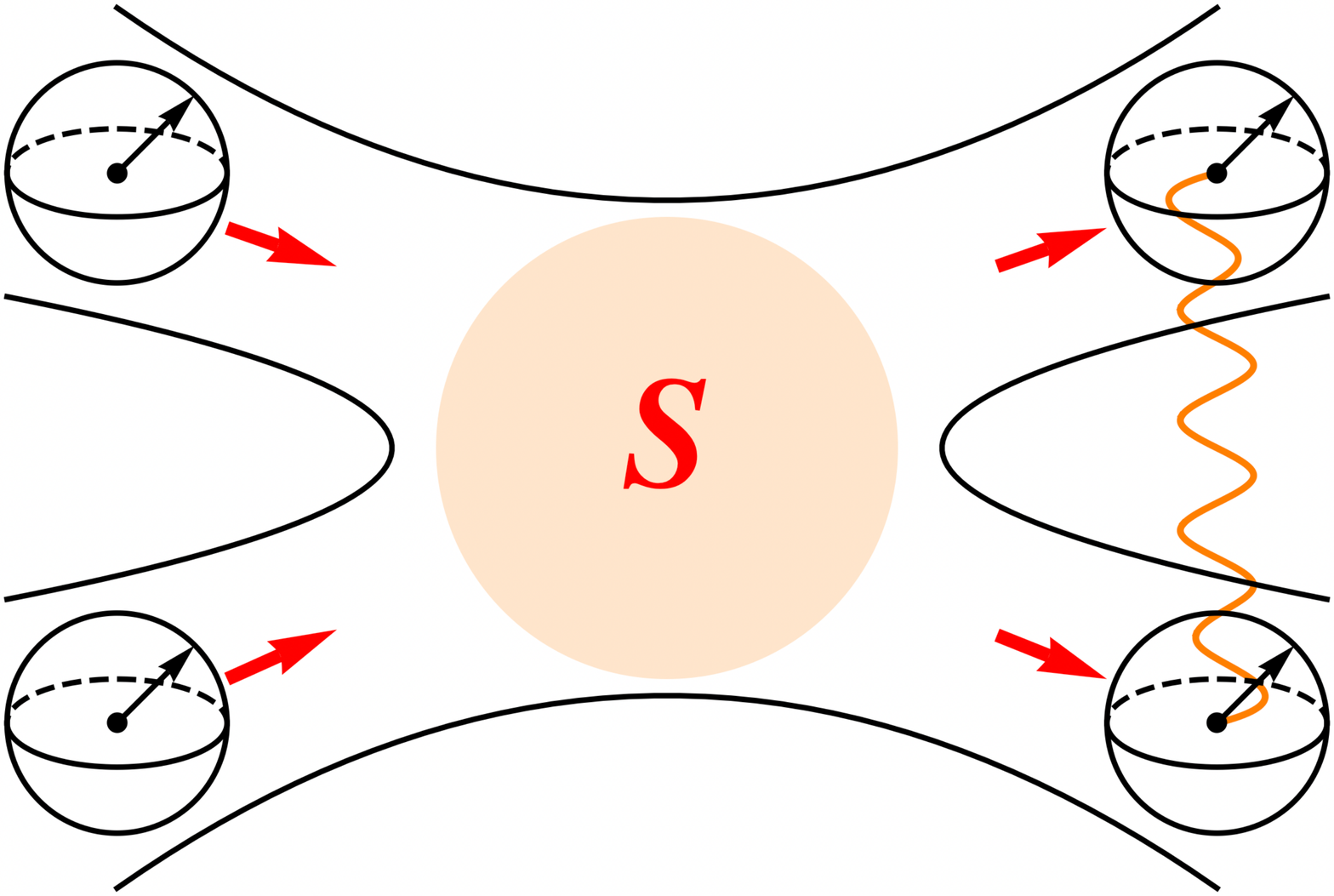}
      
  \caption{
An illustration of entanglement generation in the two-qubit scattering.
The left-hand side refers to the in-state of the scattering process,
where two unentangled qubits (denoted by the Bloch spheres) fly in from the distant past.
The light orange disk in the middle refers to the interacting stage of the scattering process,
which generates entanglement between two qubits.
The right-hand side refers to the out-state of the scattering process,
where the two qubits eventually become entangled (stressed by the orange curve) and fly out to the distant future.
}

  \label{QubitScattering}
  
    
\end{figure}

The entanglement power is a physical quantity which measures the entanglement generated by the $S$ matrix $\mathcal{S}$ in the quantum scattering between Qubit 1 and Qubit 2.
It is given by
\begin{align}
&\mathcal{E}(\mathcal{S})=\!\int\!\frac{\mathrm{d}\Omega_1}{4\pi}\!\int\!\frac{\mathrm{d}\Omega_2}{4\pi}\,\mathcal{E}_2(\mathcal{S}\ket{\text{in}}),\label{EP1}\\
&\mathcal{E}_2(\mathcal{S}\ket{\text{in}})=
1-\text{Tr}_1(\rho_1^2),
\end{align}
where $\rho_1=\text{Tr}_2(\rho_{12})$ is the reduced density matrix of Qubit 1
with $\rho_{12}=\mathcal{S}\ket{\text{in}}\bra{\text{in}}\mathcal{S}^\dagger$ being the two-qubit density matrix for the out-state $\ket{\text{out}}\equiv\mathcal{S}\ket{\text{in}}$,
and $\mathcal{E}_2(\mathcal{S}\ket{\text{in}})$ is the so-called 2-entropy for the out-state.
The integration variables $(\Omega_1,\Omega_2)\equiv(\theta_1,\phi_1,\theta_2,\phi_2)$ in Eq.~\eqref{EP1} 
are four angles which parametrize the spin orientations of the two incoming unentangled qubits in the Bloch-sphere representation
$\ket{\text{in}}=\left[\cos\frac{\theta_1}{2}\ket{0}_1+\exp(i\phi_1)\sin\frac{\theta_1}{2}\ket{1}_1\right]\otimes\left[\cos\frac{\theta_2}{2}\ket{0}_2+\exp(i\phi_2)\sin\frac{\theta_2}{2}\ket{1}_2\right]$,
with $\ket{0}_k$ and $\ket{1}_k$ being the spin-up and spin-down states of the $k$th qubit $(k=1,2)$.
Following Ref.~\cite{Beane:2018oxh}, we calculate the entanglement power in spin space.
The spatial wave function is omitted here,
as it plays no role in discussing entanglement generation in spin space and only introduces a global phase factor to the in- and out-states.
An illustration of the two-qubit scattering is shown in Fig.~\ref{QubitScattering}.

In Refs.~\cite{Beane:2018oxh,Beane:2021zvo,Liu:2022grf}, the entanglement power has been used to quantify the entanglement generation for the hadron-hadron scatterings, where the Coulomb interactions between charged hadrons are ignored for simplicity.
In contrast,
the Coulomb interaction is known to be important in low-energy nucleon-nucleus and nucleus-nucleus scatterings and cannot be omitted in order to reproduce experimental data.
However, the presence of the Coulomb interaction may complicate the mathematical definition of the $S$ matrix,
which has to be handled carefully in order to get physically meaningful results.
In this work, 
the screening method is adopted
to cut off the Coulomb interaction at large distances \cite{Taylor:2006},
based on which
 a careful treatment is derived for the $S$ matrix and the entanglement generation in elastic scatterings between charged qubits.

The $S$-wave Schr\"odinger equation for two charged particles is given by
\begin{align}
\left\{\frac{\mathrm{d}^2}{\mathrm{d}r^2}-2\mu \left[V_N(r)+\overline{V}_C(r)\right]+p^2\right\}u(r)=0,
\label{RSE}
\end{align}
with $\mu$ being the two-body reduced mass,
$V_N(r)$ being the short-range nuclear potential,
$p$ being the relative momentum,
and $u(r)$ being the radial wave function. 
In Eq.~\eqref{RSE}, 
 $\overline{V}_C(r)$ is the screened Coulomb potential given by
\begin{align}
\overline{V}_C(r) =\, & \frac{Z_1Z_2e^2}{r},\qquad r\leq r_S,\label{CE}\\
=\, & 0, \qquad\qquad\ \ \,\, r>r_S,
\end{align}
which replaces the original Coulomb potential by the screened one.
Here, 
$Z_1$ and $Z_2$ are the charge numbers of the two particles,
and $r_S$, much larger than the finite range $r_N$ of the nuclear potential $V_N(r)$, is the screening radius beyond which the Coulomb potential is turned off.
The elementary charge squared $e^2$ in Eq.~\eqref{CE} is related to the fine structure constant $\alpha$ by $\alpha=e^2/(\hbar c)$.
For $r_N\leq r\leq r_S$,
only the screened Coulomb interaction remains nonvanishing
and the radial wave function is given by
\begin{align}
&u(r)\nonumber\\
=&\,\frac{i}{2}\!\left[h^-(\eta,pr)-\exp\!\left(2i\delta^N\right) h^+(\eta,pr)\right]\label{rwfmid1}\\
\propto & \sin\!\left[pr-\eta\ln(2pr)+\sigma(\eta)+\delta^N\right]\quad \text{for large $r$}.
\label{rwfmid2}
\end{align}
Here, $h^-(\eta,\rho)$ is the incoming Coulomb-Hankel function in the $S$-wave,
$h^+(\eta,\rho)$ is the outgoing Coulomb-Hankel function in the $S$-wave,
$\eta=Z_1Z_2\alpha\mu/p$ is the dimensionless Sommerfeld parameter,
$\sigma(\eta)=\text{arg}\,\Gamma(1+i\eta)$ is the $S$-wave Coulomb phase shift,
and $\delta^N$ is the nuclear phase shift.
In Eq.~\eqref{rwfmid2},
 the asymptotic forms of the incoming and outgoing Coulomb-Hankel functions have been used to do the simplification.
For $r>r_S$, the screened Coulomb interaction vanishes
and the radial wave function is given by
\begin{align}
u(r)&=\text{const}\times\frac{i}{2}\left[h^-(0,pr)-\exp(2i\delta)h^+(0,pr)\right]\\
&\propto \sin(pr+\delta)\quad \text{for large $r$}.\label{rwfext}
\end{align}
Here, $\delta$ is the total phase shift.
Matching Eq.~\eqref{rwfext} with Eq.~\eqref{rwfmid2} continuously at $r=r_S$, one obtains
\begin{align}
\delta=\delta^N\!-\eta\ln(2pr_S)+\sigma(\eta).
\label{TPS}
\end{align}
Noticeably, the total phase shift $\delta$ depends explicitly on the screening radius $r_S$.
Rigorously speaking, it is not a well-defined physical quantity
as $r_S$ could be chosen arbitrarily.

\begin{figure}

  \includegraphics[width=\linewidth]{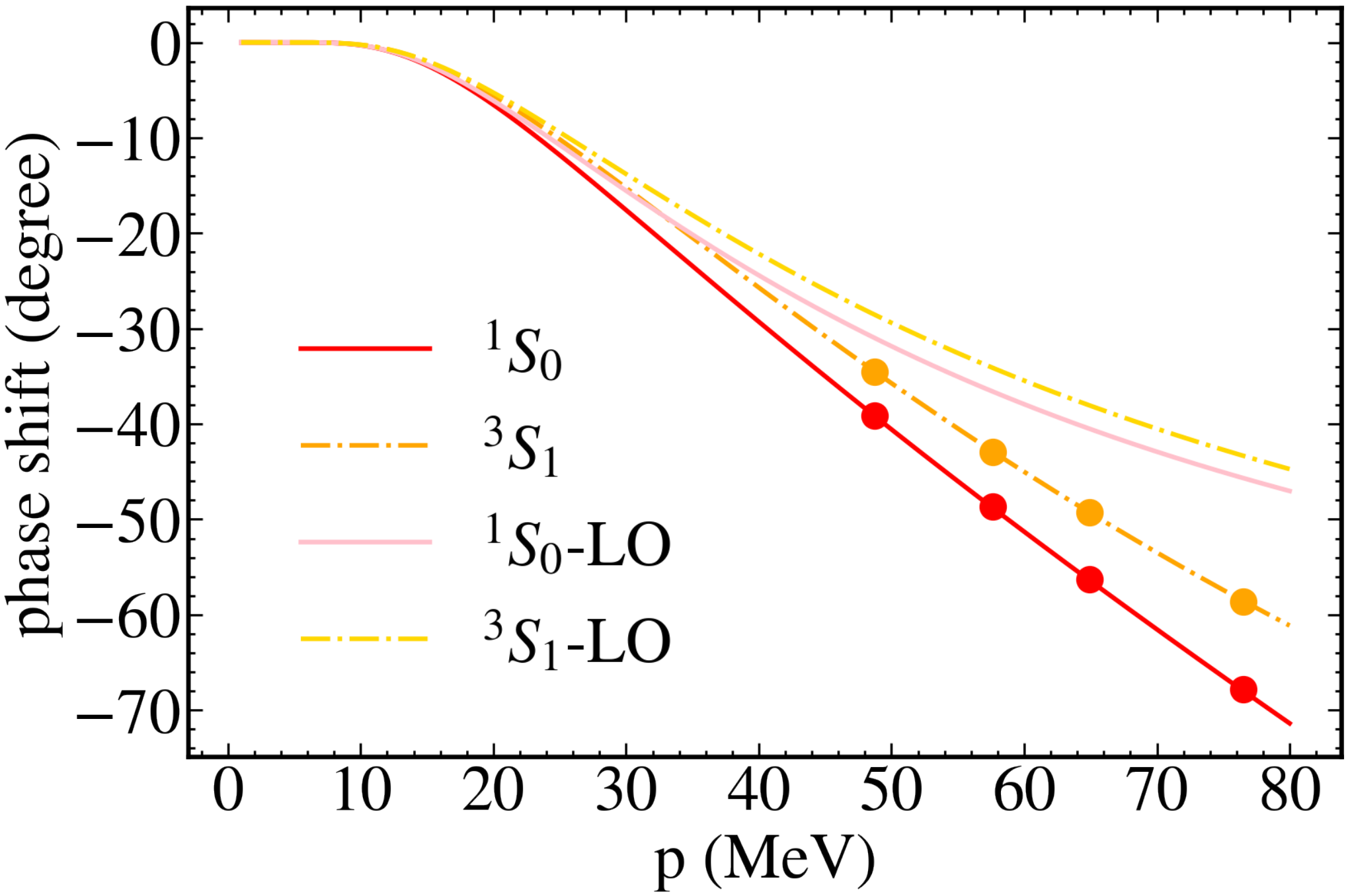}
        
  \caption{The $S$-wave phase shifts of the $p+{}^3\text{He}$ scattering are shown with respect to the relative momentum $p$.
  The red and orange points are the experimental phase shifts in the $^1S_0$ and $^3S_1$ channels from Ref.~\cite{Daniels:2010af}.
  The red solid line and the orange dash-dotted line are the phase shifts given by the Coulomb-modified EREs in Eqs.~\eqref{ERE0} and \eqref{ERE1},
  with the effective range parameters given in Sec. \ref{ETI}.
  The pink solid line and the yellow dash-dotted line are the $S$-wave phase shifts given by the Coulomb-modified EREs
  at LO, which take into consideration only the contributions from the scattering lengths.
 }

  \label{PhaseShift_p+3He}
  
    
\end{figure}

For the charged-qubit scattering,
two channels are relevant for the $S$ wave: the $^1S_0$ and $^3S_1$ channels.
The two-body scattering in each channel is described by Eq.~\eqref{RSE}, 
with $V_N(r)$ replaced by the partial-wave components of realistic nuclear potentials.
The corresponding $S$ matrix can be given by a compact form in spin space:
\begin{align}
\mathcal{S}&=\frac{1}{2}[\exp(i2\delta_1)+\exp(i2\delta_0)]\,\bm{1}_4
+\frac{1}{2}[\exp(i2\delta_1)\nonumber\\
&-\exp(i2\delta_0)]\,\text{SWAP}.
\label{SMQ}
\end{align}
Here, $\delta_0$ is the total phase shift in the $^1S_0$ channel,
$\delta_1$ is the total phase shift in the $^3S_1$ channel,
$\bm{1}_4$ is the four-dimensional unit matrix,
and $\text{SWAP}=(\bm{1}_4+\sum_{i=1}^3{\sigma}_i\otimes{\sigma}_i)/2$
is the so-called SWAP operator in quantum computation,
with $\sigma_i$ $(i=1,2,3)$ being the three Pauli matrices
and $\otimes$ being the tensor product.
In terms of Eq.~\eqref{TPS},
the total phase shifts $\delta_0$ and $\delta_1$ are given explicitly by
\begin{align}
\delta_0=\delta^N_0-\eta\ln(2pr_S)+\sigma(\eta),\label{delta0}\\
\delta_1=\delta^N_1-\eta\ln(2pr_S)+\sigma(\eta),\label{delta1}
\end{align}
where $\delta^N_0$ and $\delta^N_1$ are the nuclear phase shifts in the $^1S_0$ and $^3S_1$ channels, respectively.
The entanglement generation capacity of the $S$ matrix $\mathcal{S}$ in Eq.~\eqref{SMQ} could then be quantified by the entanglement power $\mathcal{E}(\mathcal{S})$, whose analytic expression, after some symbolic simplification, is found to be
\begin{align}
\mathcal{E}(\mathcal{S})&=\frac{1}{6}\sin^2[2(\delta_0-\delta_1)]\\
&=\frac{1}{6}\sin^2[2(\delta_0^N-\delta_1^N)].
\label{EPExp}
\end{align}
Here, Eqs.~\eqref{delta0} and \eqref{delta1} are used in deriving Eq.~\eqref{EPExp}.
Remarkably, the $r_S$ dependence in the total phase shift $\delta_0$ 
is canceled \emph{exactly} with the $r_S$ dependence in $\delta_1$,
making $\mathcal{E}(\mathcal{S})$ a well-defined physical quantity,
despite the fact that the corresponding $S$ matrix has the unpleasant dependence on $r_S$.
This $r_S$ independence could be anticipated by noting that the entanglement power is determined by the difference of the singlet and triplet phase shifts whereas the Coulomb interaction is spin independent.
Also, the Coulomb phase shifts $\sigma(\eta)$ in Eqs.~\eqref{delta0} and \eqref{delta1}
are canceled exactly in Eq.~\eqref{EPExp}.

With the nuclear phase shifts $\delta^N_0$ and $\delta^N_1$ compiled from experimental data,
the entanglement power $\mathcal{E}(\mathcal{S})$ could be determined reliably from Eq.~\eqref{EPExp}.
Making use of the Coulomb-modified effective range expansions (EREs),
the nuclear phase shifts $\delta_0^N$ and $\delta_1^N$ are parameterized up to $\mathcal{O}(p^4)$ by
\begin{align}
&C_\eta^2p\cot\delta_0^N + \gamma h(\eta)=-\frac{1}{a_0^C} +\frac{1}{2}r_0^Cp^2-\frac{1}{4}\mathcal{P}_0^Cp^4,\label{ERE0}\\
&C_\eta^2p\cot\delta_1^N + \gamma h(\eta)=-\frac{1}{a_1^C} +\frac{1}{2}r_1^Cp^2-\frac{1}{4}\mathcal{P}_1^Cp^4,\label{ERE1}
\end{align}
with $C_\eta^2=\exp(-\pi\eta)\Gamma(1+i\eta)\Gamma(1-i\eta)$,
$\gamma=2\mu Z_1Z_2\alpha=2p\eta$, $h(\eta)=\text{Re}\,\psi(i\eta)-\ln|\eta|$, and $\psi(z)$ being the digamma function.
On the right-hand sides of Eqs.~\eqref{ERE0} and \eqref{ERE1}, $a_0^C$, $r_0^C$, and $\mathcal{P}_0^C$ are the scattering length, the effective range, and the shape parameter for the $^1S_0$ channel, while $a_1^C$, $r_1^C$, and $\mathcal{P}_1^C$ are the scattering length, the effective range, and the shape parameter for the $^3S_1$ channel.
Here, the superscript ``$C$''  
stresses the presence of the Coulomb interaction.



\section{The Few-Nucleon Scatterings}
\label{NR}

In this section, we study the entanglement generation in the $S$-wave $p+{}^3\text{He}$ and $n+{}^3\text{H}$ scatterings.
The $p+{}^3\text{He}$ scattering is treated in detail in Sec.\ \ref{p3He}, 
which can be easily adapted to the $n+{}^3\text{H}$ scattering.

\subsection{The $\bm p+{}^{\bm 3}\text{He}$ scattering}
\label{p3He}

\subsubsection{Experimental inputs}
\label{ETI}

%
%
%
%
%
%

As shown in Eq.~\eqref{EPExp}, the nuclear phase shifts $\delta^N_0$ and $\delta^N_1$ in the $^1S_0$ and $^3S_1$ channels are the crucial inputs to calculate
the entanglement power $\mathcal{E}(\mathcal{S})$ in the $p+{}^{3}\text{He}$ scattering.
In this work, the phase shifts at $p\leq80$ MeV are adopted to do the calculations,
which correspond to the relative energies $\leq$ 6 MeV.
At higher relative momenta, $^3$He is likely to disintegrate and thus can no longer be treated as a qubit to a good approximation.
To estimate the values of $\delta^N_0$ and $\delta^N_1$,
we adopt the experimental extractions of effective range parameters as $a_0^C=11.1\pm0.4$ fm, $r_0^C=1.58\pm0.12$ fm, $\mathcal{P}_0^C=-4\pm0.8\ \text{fm}^3$, $a_1^C=9.04\pm0.14$ fm, $r_1^C=1.50\pm0.06$ fm, and $\mathcal{P}_1^C=0.36\pm0.32\ \text{fm}^3$ \cite{Daniels:2010af}.
The same reference also tabulates
a few experimental values of $\delta^N_0$ and $\delta^N_1$ 
at the relative momenta $p=48.71, 57.63, 64.95, 76.20$ MeV,
which are used in our calculations as well.

In Fig.~\ref{PhaseShift_p+3He},
the phase shifts from the Coulomb-modified ERE calculations are shown by the red solid line and the orange dash-dotted line for the $^1S_0$ and $^3S_1$ channels.
It is straightforward to see that they agree well with the four experimental data points (the red and orange solid points).
In the same figure, we also plot the phase shifts from the Coulomb-modified EREs at LO (the pink solid line and the yellow dash-dotted line), which consider only the scattering-length contributions on the right-hand sides of Eqs.~\eqref{ERE0} and \eqref{ERE1}.
It is found that the LO approximations can describe the $^1S_0$ and $^3S_1$ phase shifts quantitatively at $p\leq30$ MeV.  

\subsubsection{Entanglement power}

\begin{figure}

\includegraphics[width=\linewidth]{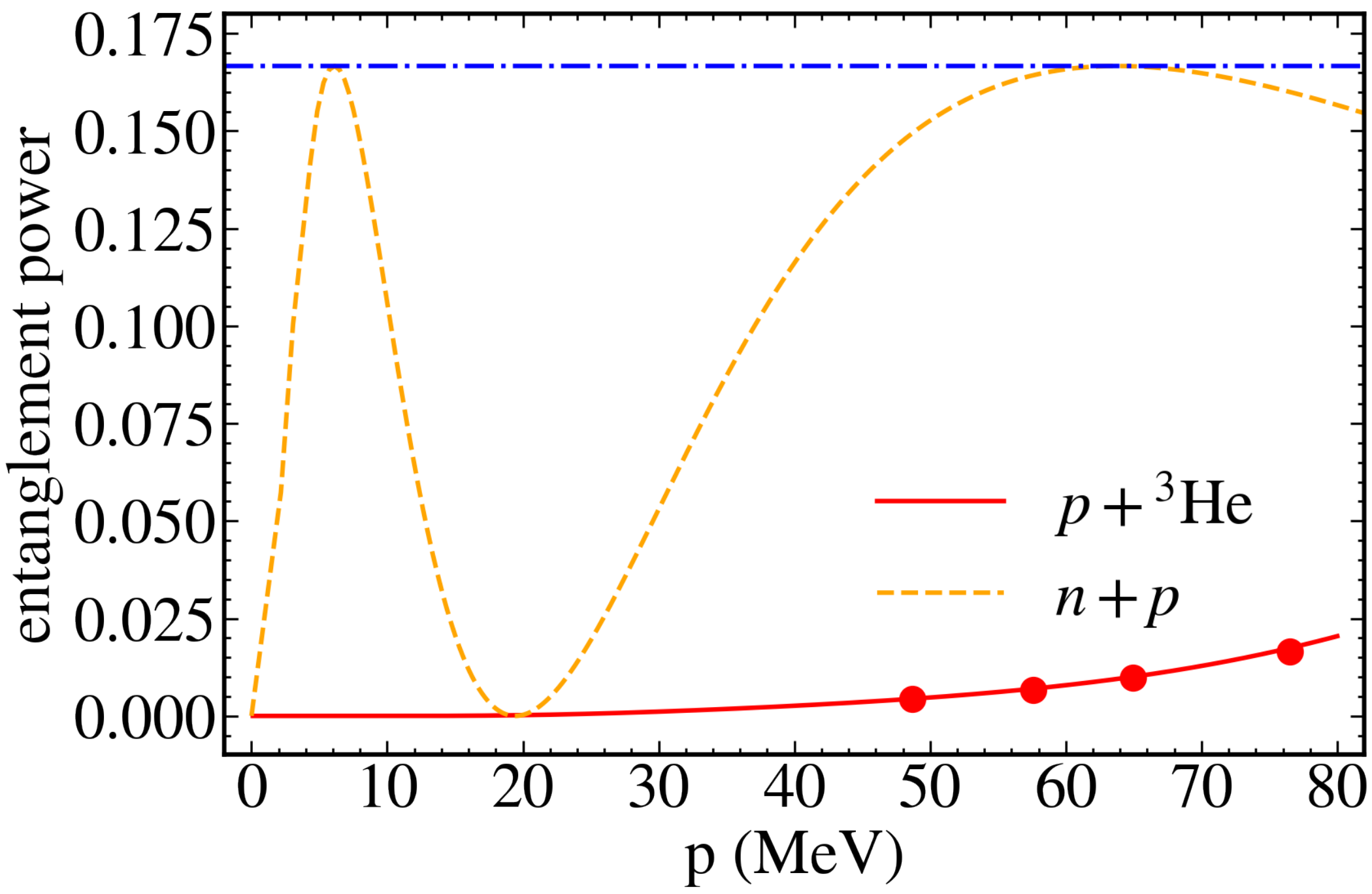}

\caption{The variations of entanglement power with respect to the relative momentum $p$.
The red solid line corresponds to the entanglement power of the $p+{}^3\text{He}$ scattering
in the $S$ wave
with the nuclear phase shifts given by the Coulomb-modified EREs in Eqs.~\eqref{ERE0} and \eqref{ERE1}.
The red solid points are the results directly obtained from the experimental phase shifts at selected relative momenta \cite{Daniels:2010af}.
The orange dashed line corresponds to
the entanglement power of the $S$-wave $n+p$ scattering
given by the Nijmegen $^1S_0$ and $^3S_1$ phase shifts.
The blue dash-dotted line gives the upper limit of the entanglement power at $\mathcal{E}(\mathcal{S})=1/6$.
}

\label{EPp3He}

\end{figure}

We study how the entanglement power $\mathcal{E}(\mathcal{S})$ varies with respect to the relative momentum $p$ in the $S$-wave $p+{}^{3}\text{He}$ scattering.
The numerical results are given by the solid red line in Fig.~\ref{EPp3He},
where $\mathcal{E}(\mathcal{S})$ is obtained via Eq.~\eqref{EPExp}
with the phase shifts $\delta^N_0$ and $\delta^N_1$ given by
the Coulomb-modified EREs in Eqs.~\eqref{ERE0} and \eqref{ERE1}.
The red solid points in Fig.~\ref{EPp3He}
are obtained directly from the experimental phase shifts at the four relative momenta mentioned, and are consistent with those given by the EREs.
For comparison, we plot in the same figure the entanglement power of the $S$-wave $n+p$ scattering (the orange dashed line)
with the $^1S_0$ and $^3S_1$ phase shifts given by the Nijmegen partial wave analysis \cite{Stoks:1993tb,NNOnline}.

As shown in Fig.~\ref{EPp3He}, 
the entanglement power of the $n+p$ scattering goes up and down and gets its maximal value of $1/6$ twice, both at $p=6.11$ and 63.76 MeV.
At these two momenta, the $S$ matrix turns the unentangled in-state $\ket{n_\uparrow}\otimes\ket{p_\downarrow}$ into the maximally entangled out-states
$(0.6362+0.3088i)\ket{n_\uparrow}\otimes\ket{p_\downarrow}+(0.3088-0.6362i)\ket{n_\downarrow}\otimes\ket{p_\uparrow}$
and $(-0.6887+0.1604i)\ket{n_\uparrow}\otimes\ket{p_\downarrow}+(-0.1604-0.6887i)\ket{n_\downarrow}\otimes\ket{p_\uparrow}$, respectively.
The maximal entanglement
of these two out-states can be verified easily by using the standard bipartite entanglement measures, among which a convenient choice could be the concurrence \cite{Hill:1997,Wootters:1998}.
In contrast, 
in the $p+{}^{3}\text{He}$ scattering,
the entanglement power $\mathcal{E}(\mathcal{S})$ increases
from 0 to around 0.02 for $p\leq80$ MeV,
which is much smaller than the $n+p$ scattering at most values of relative momenta.
In other words, the entanglement generation capacity of the effective interaction between the proton and $^3$He is much weaker than the realistic nucleon-nucleon interaction at $p\leq$ 80 MeV. 
The rise of maximums of the $n+p$ entanglement power could be understood from the fact that the $n+p$ scattering lengths 
have opposite signs for the singlet and triplet channels, 
which means that one channel is attractive whereas the other is repulsive.
In the LO effective range theory, 
the explicit form of
the $n+p$ entanglement power is found to be
\begin{align}
\mathcal{E}(\mathcal{S})=\frac{2}{3}\frac{p^2(a_0^{np}-a_1^{np})^2(1+a_0^{np}a_1^{np}p^2)^2}{(1+{a^{np}_{0}}^2p^2)^2(1+{a^{np}_{1}}^2p^2)^2},\nonumber
\end{align}
with $a_0^{np}$ and $a_1^{np}$ being the singlet and triplet scattering lengths \cite{Beane:2020wjl}.
As $a^{np}_0a^{np}_1<0$,
it is straightforward to show that
$\mathcal{E}(\mathcal{S})$ 
gets the minimum value of 0 at $p=0, \frac{1}{\sqrt{-a^{np}_0a^{np}_1}}$
and the maximal value of $1/6$ at 
\begin{align}
p=\frac{\mp a_0^{np}\pm a_1^{np}-\sqrt{{a_0^{np}}^2-6a_0^{np}a_1^{np}+{a_1^{np}}^2}}{2a_0^{np}a_1^{np}}.\nonumber
\end{align}
In contrast, both scattering lengths have the same sign in the $p+{}^3$He scattering,
thus giving rise to different low-energy behavior of $\mathcal{E}(\mathcal{S})$.
From the microscopic viewpoint,
the $p+{}^3\text{He}$ scattering is essentially a quantum four-body process involving three protons and one neutron.
One may naively expect that the few-body correlations 
 may help to promote the production of the entangled states in scattering processes,
and the entanglement generation capacity of the effective interaction between the proton and $^3$He would be stronger.
The explicit calculations show that this does not happen at least for $p\leq80$ MeV,
where the collision is dominated by the $S$-wave elastic scattering.
For higher relative momenta,
the disintegration of $^3$He is expected to be important
and the Coulomb-modified EREs in Eqs.~\eqref{ERE0} and \eqref{ERE1} are less reliable in
predicting phase shifts and entanglement power.

\subsubsection{Cluster EFT and symmetry enhancement}

We study the entanglement properties of the $p+{}^3\text{He}$ scattering
from the viewpoint of cluster EFT \cite{Hammer:2017tjm,Hammer:2019poc}.
The EFT Lagrangian is given at LO by
\begin{align}
&\mathcal{L}_\text{LO}=\mathcal{L}_\text{kin} + \mathcal{L}_\text{int},\label{Lag}\\
&\mathcal{L}_\text{kin}=\psi^\dagger\left(iD_t+\frac{\bm{D}^2}{2m}\right)\psi + \Psi^\dagger\left(iD_t+\frac{\bm{D}^2}{2M}\right)\Psi,\label{LagKin}\\
&\mathcal{L}_\text{int}= -C_S(\psi^\dagger\psi)(\Psi^\dagger\Psi)-C_T(\psi^\dagger\bm{\sigma}\psi)\!\cdot\!(\Psi^\dagger\bm{\sigma}\Psi).\label{LagInt}
\end{align}
Here, $\psi$ and $\Psi$ are the two-component spinor fields for the proton and $^3\text{He}$,
$m$ and $M$ are the masses of the proton and $^3\text{He}$, $(D_t,\bm{D})\equiv D_\mu=\partial_\mu+ie\widehat{Q}A_\mu$ is the covariant derivative,
with $A_\mu$ being the photon field and $\widehat{Q}$ being the charge operator satisfying
$\widehat{Q}\psi=\psi$ for the proton and $\widehat{Q}\Psi=2\Psi$ for $^3$He,
and $C_S$, $C_T$ are the low-energy constants (LECs) at LO.
This LO EFT is equivalent to the LO EREs.
Therefore, according to Fig.~\ref{PhaseShift_p+3He}, it can be used to describe the $S$-wave phase shifts at $p\leq30$ MeV.
The EFT Lagrangian in Eqs.~\eqref{Lag}--\eqref{LagInt} is invariant under the global transformations
$\psi\to\text{SU}(2)\psi$ and $\Psi\to\text{SU}(2)\Psi$.
Noticeably, the same $\text{SU}(2)$ transformations are taken for $\psi$ and $\Psi$.
The breakdown scale of this cluster EFT is estimated to be $\Lambda_b\sim \sqrt{2\mu_{pd} S_p}\approx80\ \text{MeV}$,
with $\mu_{pd}$ being the reduced mass of the proton-deuteron system and $S_p$ being the one-proton separation energy of $^3$He. 
The LECs $C_S$ and $C_T$ are related to the scattering lengths $a_0^C$ and $a_1^C$ by \cite{Kong:1999sf}
\begin{align}
\frac{2\pi}{\mu C_0}=\frac{1}{a_0^C}+4\alpha\mu\left[\ln\left(\frac{\pi\Lambda}{4\alpha\mu}\right)-\gamma_E\right]-\Lambda,\label{LEC1}\\
\frac{2\pi}{\mu C_1}=\frac{1}{a_1^C}+4\alpha\mu\left[\ln\left(\frac{\pi\Lambda}{4\alpha\mu}\right)-\gamma_E\right]-\Lambda,\label{LEC2}
\end{align}
with 
$C_0=C_S-3C_T$ and $C_1=C_S+C_T$ being the partial-wave LECs for the $^1S_0$ and $^3S_1$ channels,
$\mu=mM/(m+M)$ being the reduced mass of the proton and $^3$He, $\Lambda$ being the regularization scale introduced to regularize the divergent Feynman integrals with a sharp cutoff $\pi\Lambda/2$,
and $\gamma_E$ being the Euler constant.
At $\Lambda=\Lambda_b=80$ MeV, the ratio between $C_S$ and $C_T$ is found to be
$C_T/C_S=0.0468$, satisfying $C_T\ll C_S$. 
We also study how $C_T/C_S$ changes as $\Lambda$ is varied by, for example, 10\%.
It is found that $C_T/C_S=0.0641$ at $\Lambda=0.9\Lambda_b$
and $C_T/C_S=0.0366$ at $\Lambda=1.1\Lambda_b$, 
within the same order of magnitude as the aforementioned value of $C_T/C_S$ at $\Lambda=\Lambda_b$.
In Eqs.~\eqref{LEC1} and \eqref{LEC2},
the linear divergences are included explicitly in the definitions of LECs.
If they are suppressed following minimal subtraction, 
the ratio of LECs becomes $C_T/C_S=-0.0174$.


In the limit of $C_T\to 0$,
the spin-dependent vertex $-C_T(\psi^\dagger\bm{\sigma}\psi)\!\cdot\!(\Psi^\dagger\bm{\sigma}\Psi)$ in Eq.~\eqref{LagInt} vanishes.
It is straightforward to check that the global symmetry of the EFT Lagrangian is enhanced from $\text{SU}(2)$ to $\text{SU}(2)_1\otimes\text{SU}(2)_{2}$
and is invariant under the transformations
$\psi\to\text{SU}(2)_1\psi$ and $\Psi\to\text{SU}(2)_{2}\Psi$.
Here, the $\text{SU}(2)_1$ and $\text{SU}(2)_{2}$ transformations are generally different for the proton and $^3$He.
Due to the vanishing of the spin-spin vertex, 
the $^1S_0$ channel becomes indistinguishable from the $^3S_1$ channel,
as a result of which 
the singlet phase shift $\delta^N_0$ is equal to the triplet phase shift $\delta^N_1$.
According to Eq.~\eqref{EPExp},
the corresponding entanglement power $\mathcal{E}(\mathcal{S})$ equals zero exactly
for all relative momenta.
This is consistent with the relation between entanglement minimization and symmetry enhancement observed in Ref.~\cite{Beane:2018oxh}.

The emergence of the approximate $\text{SU}(2)_1\otimes\text{SU}(2)_2$ symmetry might be understood from the perspective of
the approximate Wigner SU(4) symmetry observed at the level of nucleons.
It is emphasized by Ref.~\cite{Konig:2016utl} that the
pionless EFT expanded around the unitarity limit,
which is shown to give a new systematic description of bound-state properties of 
$A\leq4$ nucleons, respects the Wigner SU(4) symmetry at LO
and thus does not have the spin-dependent nucleon-nucleon interaction.
If applicable to the $S$-wave $p+{}^3$He scattering as well,
such a theory will not distinguish between the $^1S_0$ and $^3S_1$ channels at LO
and thus gives rise to the $C_T\to0$ limit of cluster EFT automatically.
The applicability of this pionless EFT expanded around unitarity to few-nucleon scatterings has not been explored explicitly in the literature.
To some extent,
the small size of the entanglement power observed in the $S$-wave $p+{}^3$He scattering provides support for its applicability from the quantum-information perspective.

\subsection{The $\bm n+{}^{\bm 3}\text{H}$ scattering}
\label{n3H}

\begin{figure}

\includegraphics[width=\linewidth]{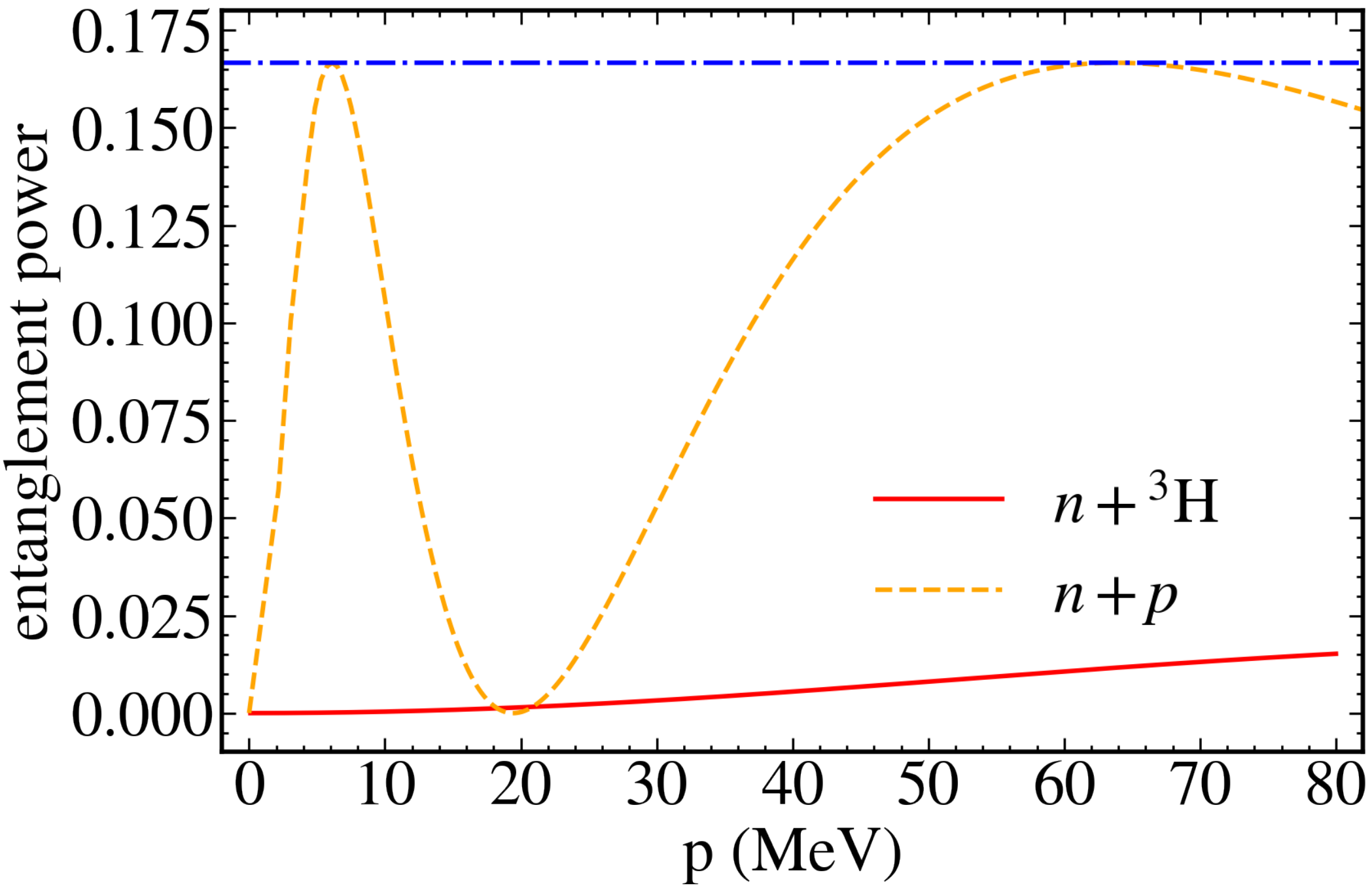}

\caption{The same as Fig.~\ref{EPp3He} except that
the red solid line gives the entanglement power for the $n+{}^3$H scattering.
}

\label{EPn3H}

\end{figure}

The above discussions on the $p+{}^3\text{He}$ scattering could be extended in parallel to the $n+{}^3$H scattering. 
As the neutron is neutral, 
the long-range Coulomb interaction does not show up in the $n+{}^3$H scattering.
The corresponding formalism of the entanglement power and cluster EFT could be obtained from Eqs.~\eqref{delta0}--\eqref{ERE1} and \eqref{Lag}--\eqref{LEC2} by taking the fine structure constant $\alpha\to0$ (i.e., the charge unit $e\to0$) effectively.
We take for the $S$-wave $n+{}^3$H scattering the following effective range parameters estimated by the latest pionless EFT calculations: 
$a_0=4.035(65)$ fm, $r_0=2.17(15)$ fm 
and $a_1=3.566(47)$ fm, $r_1=1.76(41)$ fm for the $^1S_0$ and $^3S_1$ channels \cite{Schafer:2022hzo}.\footnote{In Ref.~\cite{Schafer:2022hzo}, the standard power-counting scheme for pionless EFT is adopted, which is different from the power-counting scheme adopted by the pionless EFT expanded around unitarity \cite{Konig:2016utl}.}
With these inputs, the entanglement power $\mathcal{E}(\mathcal{S})$
as a function of the relative momentum $p$ is given 
in Fig.~\ref{EPn3H} for the $n+{}^3$H scattering.
Similarly to the $p+{}^3$He scattering,
it is found that the entanglement power $\mathcal{E}(\mathcal{S})$
is generally much smaller than the $n+p$
scattering at $p\leq80$ MeV.
In the LO cluster EFT,
whose breakdown scale is estimated to be $\Lambda_b\sim \sqrt{2\mu_{nd} S_n}\approx85\ \text{MeV}$ for the $n+{}^3$H scattering,
the ratio of the LECs $C_S$ and $C_T$ is found to be
$C_T/C_S=0.0466$ at $\Lambda=\Lambda_b$,
once again satisfying $C_T\ll C_S$.
Moreover, in the limit $C_T\to0$,
the entanglement power vanishes for all relative momenta
and the LO cluster-EFT Lagrangian has the symmetry enhancement 
from SU(2) to $\text{SU(2)}_1\otimes\text{SU(2)}_2$
in a way similar to the $p+{}^3$He scattering.

The enhanced symmetries in the $p+{}^3$He and $n+{}^3$H scatterings could also be understood by noticing that the singlet and triplet scattering lengths satisfy $a_{0}^C\sim a_1^C\sim 10$ fm and $a_0\sim a_1\sim 4$ fm for these two processes. 
From Eqs.~\eqref{LEC1} and \eqref{LEC2},
one can show that $|C_T/C_S|\propto |1/a_0^{(C)}-1/a_1^{(C)}|\ll 1$
under the condition of $a_0^{(C)}\sim a_1^{(C)}$,
which then gives the approximate $\text{SU}(2)_1\otimes\text{SU}(2)_2$ symmetry in the EFT Lagrangians.
Similarly, the small entanglement power at low energies could also be understood from $a_0^{(C)}\sim a_1^{(C)}$.
According to the LO EREs and the explicit expression of $\mathcal{E}(\mathcal{S})$ in  Eq.~\eqref{EPExp},
the property of $a_0^{(C)}\sim a_1^{(C)}$ leads naturally to $\delta_0^N\sim\delta_1^N$ and $\mathcal{E}(\mathcal{S})\propto \sin^2[2(\delta_0^N-\delta_1^N)]\sim0$ at low energies.
Moreover, 
it is interesting to explore why the approximate Wigner $\text{SU}(4)$ symmetry does not lead to a similar suppression of the entanglement in the low-energy nucleon-nucleon scattering.
The effective range parameters are estimated to be
$a_0^{np}=-23.740$ fm, $a_1^{np}=5.419$ fm, $r_0^{np}=2.77$ fm, and $r_1^{np}=1.753$ fm
 for the $n+p$ scattering \cite{Machleidt:2000ge}.
In the LO pionless EFT,
one has $C_T\propto 1/a_0^{np}-1/a_1^{np}$, and the corrections to the Wigner $\text{SU(4)}$ symmetry
 are small as long as both $a_{0}^{np}$ and $a_1^{np}$ are large.
 The approximate Wigner $\text{SU(4)}$ symmetry results directly from $a_{0,1}^{np}\gg r_{0,1}^{np}$,
 even if $a_0^{np}$ is very different from $a_1^{np}$ (with opposite signs and $|a_0^{np}/a_1^{np}|\sim 4$) \cite{Mehen:1999qs}.
 On the other hand, 
 at low energies, one has $\tan\delta_0-\tan\delta_1\sim (a_1^{np}-a_0^{np})p$.
 The large $a_1^{np}-a_0^{np}$
 thus gives rise to sizable $\delta_0-\delta_1$ in the energy range under consideration,
 which eventually leads to different behavior of $\mathcal{E}(\mathcal{S})$ in the $n+p$ scattering.

\section{Conclusions}
\label{Concl}

In this work, we study the entanglement generation capacity of effective interactions 
between nucleons and light nuclei
in the low-energy $p+{}^3\text{He}$ and $n+{}^3\text{H}$ scatterings
and compare the results to those of the realistic nucleon-nucleon interactions.
To deal with the Coulomb interaction between the proton and $^3$He properly,
the screening method is adopted to derive the 
entanglement power for charged qubits,
which, as far as we know, has not been reported explicitly in the literature.
With the derived formulas, it is found that
the entanglement power in the $p+{}^3\text{He}$ scattering
is typically much smaller than that in the $n+p$ scattering
for relative momenta $\leq 80$ MeV.
This entanglement suppression effect is found in the $n+{}^3\text{H}$ scattering as well.
Moreover,
it is accompanied by the approximate symmetry enhancement from $\text{SU}(2)$ to $\text{SU}(2)_1\otimes \text{SU}(2)_2$ in the corresponding cluster EFTs at LO,
consistent with the connection between entanglement minimization and symmetry enhancement proposed by previous studies.
Our study suggests that the entanglement generation capacities of effective interactions
between nucleons and light nuclei could be more suppressed than those of realistic nucleon-nucleon interactions at low energies.

\begin{acknowledgments}
 
This work is supported by the National Natural Science Foundation of China (Grants No.\ 11905103, No.\ 11947211, No.\ 12035011, No.\ 11975167, and No.\ 11961141003),
by the National Key R\&D Program of China (Contract No.\ 2018YFA0404403),
and by the Science and Technology Development Fund of Macau (Grant No.\ 0048/2020/A1).

\end{acknowledgments}

\end{document}